\documentclass[12pt,thmsa]{article}
\usepackage{sw20aip}

\begin{document}

\author{Emilio Santos}
\title{Can modified gravity prevent the gravitational collapse to black hole?}
\date{December 21, 2011 }
\maketitle

\begin{abstract}
It is shown that the theoretical arguments for the unavoidable collapse of
massive stars, which are valid within standard general relativity, are not
valid in extended gravity. In particular a calculation is reported of
neutron stars within a theory obtained from an Einstein-Hilbert action
involving a function of $R^{2}-1/2R_{\mu \nu }R^{\mu \nu }$ added to the
Ricci scalar $R$. The calculation suggests that within that theory there are
stable neutron stars with arbitrarily high baryon number.

\end{abstract}

\section{Introduction}

Claims for the existence of black holes rest upon observational evidence
combined with theoretical arguments. Here I analyze whether the
theoretical arguments, valid within general relativity (GR), remain valid
when GR is modified, for instance by adding other terms to the Ricci scalar
in the Einstein-Hilbert action.

Black holes have a long and interesting history\cite{Thorne}. The
realization of a Schwarzschild singularity was taken as a serious
possibility in 1930, after the work of Chandrasekhar. He proved that white
dwarf stars of high enough mass (and low enough temperature) are unstable
against gravitational collapse. As is known today those stars do not
collapse directly to black holes but suffer a supernova explosion, a core
remaining in the form of a neutron star which is believed to evolve towards
a black hole if the mass is large. The idea of collapsing stars had a strong
oposition on the part of leading astrophysicists of the time, in particular
Eddington who assumed that some effect, not yet known, would prevent the
collapse.

The next important date was 1939 when Einstein, then in Princeton, and
Oppenheimer in California wrote three important papers which I shall comment
in the following. All three papers dealt with spherically symmetric bodies,
which might be studied using standard coordinates with metric 
\begin{equation}
ds^{2}=\exp \left( \alpha \left( r\right) \right) dr^{2}+r^{2}d\Omega
^{2}-\exp \left( \beta \left( r\right) \right) dt^{2}.  \label{metric}
\end{equation}
In these coordinates the GR\ hydrostatic equilibrium equation is 
\begin{equation}
\frac{dp}{dr}+2\frac{q-p}{r}+\frac{k\left( m+4\pi r^{3}p\right) }{8\pi
r^{2}-2kmr}\left( \rho +p\right) =0, m\equiv \int_{0}^{r}4\pi
r^{2}\rho dr,  \label{1}
\end{equation}
where $k$ is $8\pi $ times the Newton constant, $r$ the radial coordinate,
and $\rho ,p$ and $q$ are the mass density, the radial pressure and the
transverse pressure, respectively. Oppenheimer and Volkoff (OV)\cite{OV}
solved eq.$\left( \ref{1}\right) $ with the assumption of local isotropy,
that is $p=q$, and the equation of state, $p=p(\rho ),$ appropriate for a
fluid of free (non-interacting) neutrons. The result was an one-parameter
family of equilibrium configurations where the total mass, $M\equiv m\left(
\infty \right) ,$ increases with the central density until a maximum about $%
0.7$ solar masses (the $OV$ limit) and then decreases. The relevant
conclusion was that there are no equilibrium configurations of free neutrons
with mass above the $OV$ limit. Later on it has been proved that for
interacting neutrons a limit still exists although it is somewhat higher,
about $2$ solar masses. Thus the question was, what happens to the core of a
white dwarf after a supernova explosion when the core has a mass above the $%
OV$ limit?. In order to answer this question Oppenheimer and Snyder\cite{OS}
studied the evolution of a spherical cloud of dust. The result was that
whatever is the initial configuration the whole cloud collapses to a black
hole. This suggests that any neutron star with a mass above the $OV$ limit
would suffer a collapse, qualitatively similar to the cloud of dust. Later
calculations have confirmed this assumption, which has lead to the current
wisdom that all observed compact (cold) objects with mass above the $OV$
limit are black holes\cite{Camenzind}.

Einstein\cite{Einstein} studied a model of star consisting of a spherical
distribution of particles each one moving in a circle around a central
point. Thus the matter in the star may be considered as a fluid with extreme
local anisotropy, that is nil radial pressure, $p=0$. If this is put in the
hydrostatic equilibrium eq.$\left( \ref{1}\right) $ we get 
\begin{equation}
q\left( r\right) =\frac{km\left( r\right) }{8\pi r-2km\left( r\right) }\rho
\left( r\right) ,  \label{2}
\end{equation}
which has a solution for any (spherical) density distribution. In fact,
given any function $\rho \left( r\right) $ we may get $m(r)$ via the second
eq.$\left( \ref{1}\right) $ whence eq.$\left( \ref{2}\right) $ gives $%
q\left( r\right) .$ Actually those functions $\rho \left( r\right) $ which
lead to a pressure violating the constraint $q\left( r\right) \leq $ $\rho
\left( r\right) $ should be excluded because this would mean that some of
the constituent particles move with a velocity greater than that of light, $c
$. On the other hand it is easy to prove that the solutions of eq.$\left( 
\ref{2}\right) $ with $q\left( r\right) <$ $\rho \left( r\right) $ are
stable against homologous collapse, which may be represented by a family of
configurations depending on a parameter $\lambda $ such that 
\[
r=\lambda x,\rho \left( \lambda x\right) =\lambda ^{-3}\rho \left( x\right)
,q\left( \lambda x\right) =\lambda ^{-3}p\left( x\right) ,m\left( \lambda
x\right) =m\left( x\right) .
\]
That is we should consider a sequence of configurations fulfilling 
\[
\frac{q\left( \lambda x\right) }{\rho \left( \lambda x\right) }=\frac{%
q\left( x\right) }{\rho \left( x\right) }=\frac{km\left( x\right) }{8\pi
\lambda x-2km\left( x\right) },
\]
with decreasing values of $\lambda .$ It is obvious that the ratio $q/\rho $
will increase indefinitely for $\lambda \rightarrow km\left( x\right) /(4\pi
x)$. Then the homologous collapse should stop before the constraint $q\left(
r\right) \leq $ $\rho \left( r\right) $ is violated. From this result
Einstein concluded that the existence of a limiting velocity, $c,$ prevents
the collapse to a Schwarzschild singularity, and he even asserted that for
any star model (i. e. not necessarily fulfilling $p=0$) this would be the
case\cite{Einstein}.

As is well known today Einstein was wrong, because in the real world
particles interact with each other and the interactions tend to produce
local isotropy. Actually between local isotropy, $p=q$, assumed by OV\cite
{OV} in their neutron star calculation and extremal local anisotropy, $p=0$,
as in Einstein model, there are intermediate cases. Indeed calculations have
been performed which prove that for some amount of local anisotropy, $q\neq p
$, the $OV$ limit may be surpassed in neutron stars\cite{Corchero}. However
it is easy to prove that stars in stable equilibrium should have local
isotropy, which in particular shows that Einstein\'{}s star model is
unstable. In fact, let us assume an equilibrium configuration with density
distribution $\rho \left( r\right) ,$ $p(r)$ and $q(r)$ being the radial and
transverse pressures respectively. Now, in the absence of external electric
or magnetic fields a fluid of neutrons with given baryon number density, $%
n\left( r\right) ,$ has the minimal energy density for an isotropic velocity
distribution, that is $p(r)=q(r).$ As a consequence if we start with an
equilibrium star configuration presenting local anisotropy and change it to
another configuration with the same $n\left( r\right) $ but local isotropy
everywhere, then the energy density decreases at every point where
previously $q\neq p.$ In summary we pass from an equilibrium configuration
with local anisotropy to another configuration, maybe out of equilibrium,
with the same baryon number but a smaller total mass. Consequently the
equilibrium star configuration is not stable.

Many authors have been reluctant to accept the existence of a spacetime
singularity in the line of Eddington and Einstein. There are good reasons
for that. In fact according to $GR$ the metric of spacetime is ruled by
second order differential equations and therefore it is natural to ask that
the metric coefficients are twice derivable everywhere, with no singularity.
The current solution to the problem is to assume that the singularity which
appears in classical $GR$ may be cured by quantum effects. But the good
solution would be that quantum effects could modify $GR$ in such a way that
stars never collapse. On this line there is a recent proposal of
``gravastars'', that is stars with a core where the vacuum, with equation of
state $p=-\rho ,$ provides a strong repulsive pressure able to stop the
collapse\cite{Mazur}. In this paper I propose a more natural solution,
namely that the correct gravitational theory is an extended GR, derived from
a modified Einstein-Hilbert action \cite{Schmidt}, and that in such gravity
theory stars never collapse. The purpose of this paper is to provide
arguments suggesting that this possibility is worth to be seriously studied.

\section{Slightly relativistic stars in f(R) gravity}

The most popular stars in Newtonian gravity are polytropes, that is stars
consisting of a fluid with equation of state 
\[
p=K\rho ^{\gamma }, 
\]
where $K$ and $\gamma $ are constant. It is well known that polytropes are
stable (unstable) if $\gamma $ is greater (smaller) than $4/3$. It is the
case that the most important types of Newtonian (or slightly relativistic)
stars where gravitational collapse is currently assumed are (massive cold)
white dwarfs and supermasive stars\cite{Shapiro} and both may be treated as
close to polytropes with $\gamma =4/3.$ Thus these stars are at the limit of
stability and it is believed that they become unstable due to GR
corrections. In fact for a slightly relativistic star eq.$\left( \ref{1}%
\right) $ may be written, assuming local isotropy, that is $p=q$, 
\begin{equation}
\frac{dp}{dr}=-\frac{k\left( m+4\pi r^{3}p\right) }{8\pi r^{2}-2kmr}(\rho
+p)\simeq -\frac{km\rho }{8\pi r^{2}}(1+\frac{4\pi r^{3}p}{m}+\frac{p}{\rho }%
+\frac{km}{4\pi r}).  \label{2a}
\end{equation}
The main term in the right side correspond to the Newtonian approximation
and the remaining 3 terms are the $GR$ corrections. It is easy to see that
every GR correction contributes to increase the gravitational field with
respect to the Newtonian one, therefore inducing unstability towards
collapse. In the following I show that corrections derived from $f(R)$
gravity have precisely the opposite effect.

$F(R)$ gravity derives from the Einstein-Hilbert action 
\begin{equation}
S=\frac{1}{2k}\int d^{4}x\sqrt{-g}\left( R+F\right) +S_{mat},  \label{3}
\end{equation}
where $R$ is the Ricci scalar and $F\left( R\right) $ is an arbitrary
function or $R$, which I shall assume twice derivable. The action eq.$\left( 
\ref{3}\right) $ gives the field equation\cite{Faraoni} 
\begin{equation}
(1+F_{1}\left( R\right) )R_{\mu \nu }-\frac{1}{2}g_{\mu \nu }\left(
R+F\left( R\right) \right) =\nabla _{\mu }\nabla _{\nu }F_{1}-g_{\mu \nu
} F_{1}+kT_{\mu \nu },  \label{4}
\end{equation}
where $F_{1}\equiv dF/dR$ and $R_{\mu \nu }$ ($T_{\mu \nu })$ is the Ricci
(stress-energy) tensor. The divergence of eq.$\left( \ref{4}\right) $ leads
to 
\begin{equation}
k\nabla ^{\mu }T_{\mu \nu }=R_{\mu \nu }\nabla ^{\mu }F_{1}\left( R\right)
+\left( \nabla _{\nu } - \nabla _{\nu }\right) F_{1}\left( R\right) .
\label{4a}
\end{equation}

For a static spherical body in the metric eq.$(\ref{metric})$ only the
component with $\nu =1$ of the tensor eq.$\left( \ref{4a}\right) $ gives a
nontrivial result, that is (for local isotropy $p=q$) 
\begin{equation}
\frac{dp}{dr}+\frac{1}{2}\beta ^{\prime }(\rho +p)=k^{-1}F_{2}\left(
R\right) R^{\prime }\left( R_{1}^{1}+\frac{1}{2}\beta ^{\prime \prime }-%
\frac{1}{2}\alpha ^{\prime \prime }-\frac{2}{r^{2}}\right) ,  \label{5a}
\end{equation}
where $F_{2}\equiv d^{2}F/dR^{2},R^{\prime }=dR/dr,\beta ^{\prime }\equiv
d\beta /dr,\beta ^{\prime \prime }\equiv d^{2}\beta /dr^{2},\alpha ^{\prime
\prime }\equiv d^{2}\alpha /dr^{2}.$ Now let us assume that the deviation
from GR is not large, that is the right side of eq.$\left( \ref{5a}\right) $
is in some sense small (note that the left side equated to zero just gives
the GR eq.$\left( \ref{1}\right) ).$ Then for a slightly relativistic star
it is consistent to neglect the terms $\alpha ^{\prime \prime }$ and $\beta
^{\prime \prime }$ in the right hand side and we get 
\[
\frac{dp}{dr}+\frac{1}{2}\beta ^{\prime }(\rho +p)\simeq -\frac{1}{2}\beta
^{\prime }(\rho +p)+k^{-1}F_{2}\left( R\right) R^{\prime }\left( R_{1}^{1}-%
\frac{2}{r^{2}}\right) .
\]
Similarly we have, from Einstein\'{}s equation, 
\[
R_{1}^{1}=\frac{1}{2}k\left( \rho -p\right) \simeq \frac{1}{2}k\rho ,
\]
where I have taken into account that $p<<\rho $ for an almost Newtonian
star. But in addition we have 
\[
\frac{4}{3}\pi \rho r^{3}<m,\frac{km}{r^{3}}<<\frac{2}{r^{2}},
\]
the former inequality being a consequence of the fact that the density
decreases with increasing radius and the latter is true for slightly
relativistic stars where $km<<4\pi r$. Thus we may neglect the term with $%
R_{1}^{1}$ whence, taking eq.$\left( \ref{2a}\right) $ into account eq.$%
\left( \ref{5a}\right) ,$ gives finally the following effective
gravitational field

\begin{equation}
g\equiv \frac{1}{\rho }\frac{dp}{dr}\simeq -\frac{km}{8\pi r^{2}}(1+\frac{%
4\pi r^{3}p}{m}+\frac{p}{\rho }+\frac{km}{4\pi r})-2F_{2}\left( R\right) 
\frac{1}{r^{2}\rho }\frac{d\rho }{dr}.  \label{6}
\end{equation}

The relevant result is that the latter term, which is the correction derived
from $f(R)$ gravity, is always positive because $d\rho /dr<0$ and the
inequality $F_{2}\left( \rho \right) >0$ holds true for any viable $f(R)$
theory\cite{Faraoni}. In conclusion any modification of GR via f(R) theory
tends to compensate for the unstability created by GR in slightly
relativistic stars. Whether this is enough to prevent the gravitational
collapse of white dwarfs or supermassive stars may depend on the form and
size of the function $F(R)$, but I will not pursue the study here. In
summary, although I cannot claim that $f(R)$ theory always prevents the
gravitational collapse of white dwarfs and supermassive stars, it is true
that their collapse is not unavoidable in $f(R)$ theory, at a difference
with standard $GR.$

\section{Neutron stars in quartic gravity}

Our second example will be the case of neutron stars in quartic gravity,
where the Lagrangian $F$ of eq.$\left( \ref{3}\right) $ is\cite{SantosPRD}  
\begin{equation}
F=aR^{2}+bR_{\mu \nu }R^{\mu \nu },a=-b/2=1\textrm{km}^{2}.  \label{8}
\end{equation}
The gravity theory resulting from this choice is not viable for, at least,
two reasons. Firstly it contradicts Solar System and terrestrial
observations or experiments, which require $\sqrt{a}$ and $\sqrt{-b}$ not
greater than a few millimeters. Secondly the weak field limit of the theory
presents ghosts. The difficulties may be solved if we assume that the theory
resting upon eqs.$\left( \ref{3}\right) $ and $\left( \ref{8}\right) $ is an
approximation, valid for fields of the order of those appearing inside or
near neutron stars, of a theory giving a negligible correction to GR for
weak fields. For instance, we might assume that eq.$\left( \ref{8}\right) $
is an approximation of the theory resting upon the Lagrangian 
\begin{equation}
F=aR^{2}+bR_{\mu \nu }R^{\mu \nu }-c\log \left( 1+(a/c)R^{2}+(b/c)R_{\mu \nu
}R^{\mu \nu }\right) .  \label{9}
\end{equation}
In particular if we choose $c\sim 1/(10^{20}$ km$^{2})$ eq.$\left( \ref{9}%
\right) $ becomes 
\begin{eqnarray}
F &\simeq &\frac{1}{2c}\left( aR^{2}+bR_{\mu \nu }R^{\mu \nu }\right) ^{2}%
\sim 10^{-32}R\textrm{ for the Solar System,}  \nonumber \\
F &\simeq &aR^{2}+bR_{\mu \nu }R^{\mu \nu }\sim 10^{-3}R\textrm{ for
neutron stars,}  \label{10}
\end{eqnarray}
and the relative error due to the terms neglected in going from eq.$\left( 
\ref{9}\right) $ to eq.$\left( \ref{10}\right) $ is smaller than $10^{-12}$
in both cases. I have taken into account that $R^{2}\sim R_{\mu \nu }R^{\mu
\nu }\sim (k\rho )^{2}$ and that typical densities are $\rho \sim 10^{4}$
kg/m$^{3}$ in the Solar System and $\rho \sim 10^{18}$ kg/m$^{3}$ in neutron
stars. I conclude that, in this example, the correction to $GR$ would be
negligible for the Solar System (and even more so for the cosmology at
present time), but quite important for neutron stars. In any case it is
irrelevant for the purposes of this paper the question whether the theory is
viable or not. Indeed I do not pretend to study neutron stars with the
``correct'' gravity theory which is unknown, but to provide a counterexample
to the conjecture that stable neutron stars cannot exist with more than
twice the baryon number of the Sun. The counterexample proves that such a
limit does not hold true in extended gravity theories.

In the following I report on the results obtained from the calculation of
noninteracting neutron stars using the gravity theory defined in eq.$\left( 
\ref{8}\right) .$ The field equation is\cite{Turner} 
\begin{eqnarray}
&&G_{\mu \nu }+a\left[ -GG_{\mu \nu }+\frac{1}{4}g_{\mu \nu
}(G^{2}-G_{\lambda \sigma }G^{\lambda \sigma })-G_{\mu }^{\sigma }G_{\sigma
\nu }\right]  \nonumber \\
&&+\frac{1}{2}a\left[  G_{\mu \nu }-\nabla _{\sigma }\nabla _{\nu
}G_{\mu }^{\sigma }-\nabla _{\sigma }\nabla _{\mu }G_{\nu }^{\sigma }+
G_{\mu \nu }\right] =kT_{\mu \nu },  \label{11}
\end{eqnarray}
written in terms of the Einstein tensor $G_{\mu \nu }=R_{\mu \nu }-\frac{1}{2%
}g_{\mu \nu }R$ rather then the Ricci tensor $R_{\mu \nu }$. When the
components of $G_{\mu \nu }$ are written in terms of the parameters $\alpha $
and $\beta $ of the metric and their derivatives\cite{Synge} we get a system
of 3 independent fourth order differential equations (one of the four
components of the tensor eq.$\left( \ref{11}\right) $ is not independent
because in spherical symmetry $G_{22}=G_{33}.)$ The constraints that $\alpha 
$ and $\beta $ are finite at the origin and go to zero at infinity fix all
the initial or boundary conditions except one, so that the solution of the 3
(coupled, nonlinear) ordinary differential equations provides a
one-parameter family depending of the central value of the component $%
G_{1}^{1}$. For simplicity in the calculation I have used as equation of
state the following 
\begin{equation}
\rho _{mat}=3p_{mat}+Cp_{mat}^{3/5}, C=8.17\times 10^{7}\textrm{ kg}%
^{2/5}\textrm{m}^{-6/5},  \label{eos}
\end{equation}
which has the correct behaviour for a fluid of free neutrons both at high
and low densities. The constant $C$ is so chosen that a general relativistic
calculation gives the same result as the original Oppenheimer and Volkoff%
\cite{OV} \ for the maximum mass stable star (see Table 2 below).

A relevant quantity is the baryon number of the star, $N$, which may be
calculated from the baryon number density $n(r)$ via 
\begin{equation}
N=\int_{0}^{R}\frac{n(r)r^{2}dr}{\sqrt{1-km(r)/(4\pi r)}},  \label{N}
\end{equation}
The number density $n$ may be related to the mass density and the pressure
from the solution of the equation 
\[
p_{mat}=n\frac{d\rho _{mat}}{dn}-\rho _{mat}, 
\]
which follows from the first law of thermodynamics. Inserting eq.$\left( \ref
{eos}\right) $ here we get a differential equation which may be easily
solved with the condition $\rho _{mat}/n\rightarrow \mu $ for $\rho
\rightarrow 0$, $\mu $ being the neutron mass. I get 
\begin{equation}
n=C^{5/8}p_{mat}^{3/5}\left( 4p_{mat}^{2/5}+C\right) ^{3/8}.  \label{n}
\end{equation}

The details of the calculation may be seen elsewhere\cite{Santos}. In 
Table 1 I give the results obtained. For comparison in \emph{%
Table 2} the results of a calculation within GR are presented.

\textbf{Table 1}. \emph{Calculation within extended gravity theory. }The
parameter $k$ is $8\pi $ times Newton\'{}s constant. The central matter
density, $\rho _{mat}(0),$ is in units $\rho _{c}\equiv 7.2\times 10^{18}$
kg/m$^{3}$ and similar units are used for the products of $k^{-1}$ times the
components of the Einstein tensor, $G_{\mu }^{\nu },$ at the star center.
Central baryon number density, $n(0)$, is in units $\rho _{c}/\mu ,$ $\mu $
being the neutron mass. The star radius, $R,$ is in kilometers, the mass, $M,
$ in solar masses and the baryon number, $N,$ in units of solar baryon
number. I report also the dimensionless fractional surface red shift, $%
\Delta \lambda /\lambda =1/\sqrt{1-2M/R}-1,$ and the percent binding energy, 
$BE=100(N-M)/N$. An expressions like $1.6E2$ means $1.6\times 10^{2}$

$
\begin{array}{llllllll}
k^{-1}G_{1}^{1}\left( 0\right) & 0.01 & 0.1 & 1 & 10 & 100 & 1000 & 10000 \\ 
k^{-1}G_{4}^{4}\left( 0\right) & 0.18 & 0.82 & 4.5 & 34 & 3.1E2 & 3.0E3 & 
3.0E5 \\ 
\rho _{mat}\left( 0\right) & 1.6E2 & 2.5E3 & 4.3E4 & 7.8E5 & 1.6E7 & 3.2E8 & 
6.3E9 \\ 
n\left( 0\right) & 56 & 4.5E2 & 3.7E3 & 3.3E4 & 3.2E5 & 3.0E6 & 2.8E7 \\ 
R & 10.7 & 6.7 & 4.0 & 2.7 & 2.1 & 2.2 & 2.2 \\ 
M & 0.67 & 0.80 & 0.60 & 0.39 & 0.264 & 0.268 & 0.292 \\ 
N & 0.73 & 0.94 & 1.03 & 1.13 & 1.44 & 2.00 & 2.63 \\ 
BE & 8.9\% & 15\% & 41\% & 65\% & 82\% & 87\% & 89\% \\ 
\Delta \lambda /\lambda & 0.106 & 0.22 & 0.34 & 0.31 & 0.23 & 0.23 & 0.26
\end{array}
$

\textbf{Table 2.} \emph{Calculation within general relativity. }The
meaning of all symbols is as in Table 1, but here the matter pressure and
density equal $k^{-1}$ times the appropriate components of Einstein\'{}s
tensor.

$
\begin{array}{llllllll}
k^{-1}G_{1}^{1}\left( 0\right) =p(0) & 0.01 & 0.04 & 0.1 & 1 & 10 & 100 & 
1000 \\ 
-k^{-1}G_{4}^{4}\left( 0\right) =\rho \left( 0\right)  & 0.18 & 0.46 & 0.89
& 5.3 & 39 & 3.4\times 10^{2} & 3.1\times 10^{3} \\ 
R & 11.9 & 9.6 & 8.3 & 5.8 & 5.2 & 6.6 & 6.6 \\ 
M & 0.67 & 0.72 & 0.70 & 0.55 & 0.39 & 0.40 & 0.44 \\ 
N & 0.69 & 0.74 & 0.73 & 0.55 & 0.36 & 0.37 & 0.42 \\ 
BE & 2.9\% & 3.4\% & 3.4\% & -0.73\% & -8.1\% & -8.0\% & -5.7\% \\ 
\Delta \lambda /\lambda  & 0.094 & 0.13 & 0.15 & 0.18 & 0.13 & 0.11 & 0.12
\end{array}
$

\section{Discusion}

The most dramatic difference between the calculations within extended
gravity and within GR is in the variation of the baryon number with the
central density. In GR there is a maximum about $N=0.74$ times the baryon
number of the Sun for some central density, $\rho \left( 0\right) ,$ and
after that it decreases with increasing $\rho \left( 0\right) $ (which gives
the OV limit for the baryon number). In our calculation with modified
gravity $N$ increases monotonically with the central density, which suggets
that there is no limit. Of course for the extremely high densities which
appear in \emph{Table 1} the individual neutrons would not exist and a
fluid of different particles would appear, but we have used the same
equation of state (of free neutrons) throughout. It is unlikely that a
change of the equation of state could produce a very different qualitative
behaviour, taking into account that for high density the equation should
have the form $p\simeq \rho /3$ in any case. The binding energy is also
dramatically different in both calculations. Finally the gravitational
surface redshift surpases $0.3$ whilst in GR it is always smaller than $0.2$%
, which seems to agree better with observations\cite{Camenzind}. The
comparison with observations is however not too relevant because the results
do not take into account the interaction between neutrons.

If the ``correct'' gravity theory gives results qualitatively similar to
those in \emph{Table 1}, the evolution of any massive star would end in a
compact star with relatively small mass but a very large baryon number. The
essential conclusion of the paper is that the theoretical arguments
supporting the existence of black holes, which rest upon standard GR, may be
no longer valid in extended gravity theories. If it turns out that GR must
be modified (possibly by quantum effects), then it might be the case that
the alledged observational evidence for black holes should be revised.

\end{document}